\documentclass[aps,floatfix,twocolumn]{revtex4}
\usepackage{graphics,epsfig}
\usepackage{graphicx}

\begin{document}

\title{Integration of the Friedmann equation for universes of arbitrary complexity.}
\author{Kayll Lake \cite{email}}
\affiliation{Department of Physics, Queen's University, Kingston,
Ontario, Canada, K7L 3N6 }
\date{\today}
\begin{abstract}
An explicit and complete set of constants of the motion are
constructed algorithmically for
Friedmann-Lema\^{i}tre-Robertson-Walker (FLRW) models consisting
of an arbitrary number of non-interacting species, each with a
constant ratio of pressure to density. The inheritance of
constants of the motion from simpler models as more species are
added is stressed. It is then argued that all FLRW models admit a
unique candidate for a gravitational epoch function - a function
which gives a global time-orientation without reference to
observers. The same relations that lead to the construction of
constants of the motion allow an explicit evaluation of this
function. In the simplest of all models, the $\Lambda$CDM model,
it is shown that the epoch function exists for all models with
$\Lambda > 0$, but for almost no models with $\Lambda < 0$.
\end{abstract}
\maketitle

\section{Introduction}

Recent cosmological observations \cite {recent} continue to
provide strong evidence that the $\Lambda$CDM model is a
reasonable approximation to the recent history of our universe.
However, the true nature of the dominant term - the dark energy -
remains one of the foremost problems in all of science today
\cite{dark}. With the emergence of at least three major players in
current cosmology; the cosmological constant $\Lambda$ (or,
perhaps, a refinement), spatial curvature $k$ and the matter $M$
(dominated by the yet to be understood dark matter), the view that
the history of our universe is best represented by a trajectory in
phase space has become commonplace. Indeed, many (but notably not
all \cite{john}) analyses of the observations are now filtered
through the $\Lambda$CDM model and reported by way of our current
position in the $\Omega_{\Lambda}$ - $\Omega_{M}$ plane. Here we
use this phase space approach to completely solve the central
problem in classical theoretical cosmology, the integration of the
Friedmann equation. This program is completed for all models
consisting of an arbitrary number of non-interacting species and
is accomplished through the algorithmic construction of a complete
set of constants of the motion. These constants are to be
considered as the fundamental characteristics of the model
universes. In this approach the universe is viewed as a whole
without reference to any particular class of observer. The
question then naturally arises as to whether or not events in a
given universe can be ordered globally without reference to any
particular class of observer. This question is also explored by
way of the construction of a gravitational epoch function. It is
shown that such a function need not exist, but all observations
strongly suggest that we live in a universe for which it does.

\section{The Friedmann equation}
We use the following notation: $a(t)$ is the (dimensionless) scale
factor and $t$ the proper time of comoving streamlines. $H \equiv
\dot{a}/a$, $q\equiv -\ddot{a}a/\dot{a}^2$, $^{.} \equiv d/dt$,
$k$ is the spatial curvature (scaled to $\pm 1$ for $k\neq0$ and
in units of $L^{-2}$ ($L$ signifies length)), $\rho$ is the energy
density and $p$ the isotropic pressure. $\Lambda$ is the (fixed)
effective cosmological constant and $_{0}$ signifies current
values. In currently popular notation, the Friedmann equation is
given by
\begin{equation}
\Omega+\Omega_{\Lambda}+\Omega_{k}=1 \label{omegatotal}
\end{equation}
where
\begin{equation}
\Omega_{\Lambda} \equiv \frac{c^2 \Lambda}{3 H^2},\;\;\;\Omega_{k}
\equiv -\frac{c^2 k}{H^2 a^2} \label{omegas}
\end{equation}
and
\begin{equation}
\Omega\equiv\frac{8 \pi G \rho}{3 H^2 c^2}. \label{omegarho}
\end{equation}

Consider an arbitrary number of non-interacting separately
conserved species so that
\begin{equation}
\rho=\sum_{i}\rho_{i} \label{rho}
\end{equation}
with each species characterized by
\begin{equation}
p_{i}=w_{i}\rho_{i} \label{barotropic}
\end{equation}
where $w_{i}$ is a constant and distinct for each species. (In the
real universe different species interact.  The principal
assumption made here is that these interactions are not an
important effect.) The conservation equation now gives us
\begin{equation}
\frac{8 \pi G \rho_{i}}{3 c^2}
=\frac{\mathcal{C}_{i}}{a^{3(1+w_{i})}} \equiv
H^2\Omega_{i}\label{Cconstant}
\end{equation}
where for each species $\mathcal{C}_{i}$ is a constant. We can
formally incorporate $\Lambda$ and $k$ into the sum (\ref{rho}) by
writing $w_{\Lambda}=-1$ so that $\mathcal{C}_{\Lambda}=c^2
\Lambda/3$ and $w_{k}=-1/3$ so that $\mathcal{C}_{k}= -c^2 k$ and
write (\ref{omegatotal}) in the form
\begin{equation}
\sum_{i}\Omega_{i}=1 \label{omegasum}
\end{equation}
where the sum it is now over all ``species". Since $w_{i}$ is
assumed distinct for each species, there is no way to distinguish,
say, Lovelock \cite{lovelock} (\textit{i.e.} geometric) and vacuum
contributions to $\Lambda$. However, one can introduce an
arbitrary number of separate species about the ``phantom divide"
at $w=-1$.

\begin{figure}[ht]
\epsfig{file=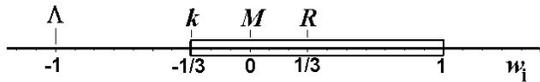,height=0.5in,width=3in,angle=0}
\caption{\label{fig1}The range in $w_{i}$ is shown (in the box)
for which the species satisfies the classical energy conditions.
Here $M$ stands for dust ($w_{M}=0$) and $R$ for radiation
($w_{R}=1/3$).}
\end{figure}

If $\mathcal{C}_{i}>0$ for species $i$ it follows that all the
classical energy conditions hold \cite{energy} for that species as
long as
\begin{equation}
-\frac{1}{3} \leq w_{i} \leq 1.\label{energy}
\end{equation}
The situation is summarized for convenience in Fig \ref{fig1}.

It follows from (\ref{Cconstant}) that
\begin{equation}
2q-1=3\sum_{i}w_{i}\Omega_{i}\label{q}
\end{equation}
and writing $1+z \equiv  \frac{a_{0}}{a}$ \cite{nophotons}, it
also follows that
\begin{equation}
\Omega_{i}=\frac{(1+z)^{3(1+w_{i})}\Omega_{io}}{\sum_{j}\Omega_{jo}(1+z)^{3(1+w_{j})}}.\label{OmegaZ}
\end{equation}
As $1+z \rightarrow \infty$ (Big Bang or Big Crunch) it follows
that $\Omega_{i} \rightarrow 0$ for all species except that with
the largest $w_{i}>-1$. Then $\Omega_{i} \rightarrow 1$. As $1+z
\rightarrow 0$, $\Omega_{i} \rightarrow 0$ for all species except
that with the smallest $w_{i}\leq-1$. Then $\Omega_{i} \rightarrow
1$. From (\ref{q}) and (\ref{OmegaZ}) it follows that the
evolution of the universes considered are governed by the
autonomous (but non-linear) system \cite{dynamical}
\begin{equation}
\Omega_{i}^{'}=\Omega_{i}((2q-1)-3w_{i})\label{system}
\end{equation}
where $^{'}\equiv -(1+z)\frac{d}{d(1+z)}$.

\section{Solving the Friedmann equation}

We solve the system (\ref{system}) algorithmically by way of the
construction of a complete set of constants of the motion.
Consider all species or any subset thereof. To distinguish this
latter possibility we replace $i$ by $\alpha$. A constant of the
motion, defined globally over the history of a universe, and not
with respect to any particular observer, is defined here to be a
function
\begin{equation}
\mathcal{F}=\mathcal{F}(\Omega_{\alpha})\label{function}
\end{equation}
for which
\begin{equation}
\mathcal{F}^{\;'}=0. \label{constant}
\end{equation}
Consider the product
\begin{equation}
\mathcal{F}=
\prod_{\alpha}\Omega_{\alpha}^{v_{\alpha}}\label{product}
\end{equation}
where the exponents $v_{\alpha}$ are constants. Note that we do
not make use of (\ref{omegasum}) at this stage. For
(\ref{constant}) to hold for (\ref{product}), it follows from
(\ref{system}) that
\begin{equation}
(2q-1)V=3W\label{constantmotion}
\end{equation}
where
\begin{equation}
V\equiv\sum_{\alpha}v_{\alpha},\;\;\;
W\equiv\sum_{\alpha}w_{\alpha}v_{\alpha}.\label{VW}
\end{equation}
If $V \neq 0$ then according to (\ref{constantmotion}) $q$ itself
is a constant and so from (\ref{system}) for each species the
evolution is given by
\begin{equation}
\Omega_{i}= \Omega_{io} (1+z)^{1-2q+3w_{i}}. \label{qconst}
\end{equation}
We do not consider this case further here. If $q$ is not constant
then for $\mathcal{F}$ of the form (\ref{product}) to be a
constant of the motion it follows from (\ref{constantmotion}) that
\begin{equation}
V=W=0 \label{vwconst}
\end{equation}
and the constant itself then reduces to
\begin{equation}
\mathcal{F}=\prod_{\alpha} \mathcal{C}_{\alpha}^{v_{\alpha}}.
\label{constantC}
\end{equation}

Sums, differences, products and quotients of constants of the
motion are of course constants as well and so the range in
$\alpha$ must be examined. First consider two species, say
$\alpha=a, b$. It then follows form (\ref{vwconst}) that
$w_{a}=w_{b}$ which is not possible as we require that the species
be distinguished by $w$. Next suppose that $\alpha$ ranges over
more than three species. It now follows from (\ref{vwconst}) that
any such constant can be reduced to products and quotients of
constants involving only three distinct species. As a result we
need consider only those constants $\mathcal{F}$ constructed from
three distinct species which we now do.

The trivial case is that of a universe for which there are only
three species, say $a, b,$ and $c$. Then
$\mathcal{F}=\Omega_{a}^{v_{a}}\Omega_{b}^{v_{b}}\Omega_{c}^{v_{c}}$
and since we can always choose $v_{a}$, (\ref{vwconst}) reduces to
two equations in two unknowns. The constant of motion is then
determined up to the choice in $v_{a}$. For example, with dust we
have the species $\Lambda, M$ and $k$ (assumed $\neq 0$
\cite{two}, the $\Lambda$CDM model) and it follows immediately
that \cite{alpha}
\begin{equation}
\mathcal{F}=\frac{\Omega_{\Lambda}\Omega_{M}^2}{\Omega_{k}^3}.
\label{dust}
\end{equation}
We can now substitute for $\Omega_{k}$ from (\ref{omegasum}).

More generally, consider $n$ species, $n>3$. If, say, species $a$
and $b$ are chosen to construct $\mathcal{F}$ then there are $n-2$
choices for the remaining species, the number of independent
constants of the motion. Further, for example, there are
$(n-2)(n-3)/2$ dependent constants of motion that follow which
involve $a$ but not $b$. In any event, as with the case of only
three species, since we can always specify one exponent,
(\ref{vwconst}) always reduces to two equations in two unknowns.

As the foregoing algorithm makes clear, if species are added to a
model, but none taken away, the constants of motion are
\textit{inherited} from the simpler model. For example,
irrespective of what we add to the $\Lambda$CDM model,
$\mathcal{F}$, given by (\ref{dust}), remains a constant of the
motion.

Now consider evolution in a three dimensional subspace, say
$\Omega_{a}, \Omega_{b}, \Omega_{c}$. We can construct the
constant $\mathcal{F}_{1}=\Omega_{a}\Omega_{b}^{v_{b}}
\Omega_{c}^{v_{c}}$, and the constant
$\mathcal{F}_{2}=\Omega_{a}^{\bar{v}_{a}}\Omega_{b}^{\bar{v}_{b}}
\Omega_{d}$ where $d$ is any species other than $a, b$ or $c$.
Then with (\ref{omegasum}) we replace $\Omega_{d}$. All species
other than $a,b$ or $c$ that now enter are replace with the aide
of constants constructed from that species and any two of $a,b$
and $c$. In the subspace then we have two independent relations
that relate $\Omega_{a}, \Omega_{b}, \Omega_{c}$. The intersection
of these surfaces in the subspace defines the evolution trajectory
of the associated universe in that subspace.

To amplify the foregoing, consider, for example,  dust and
radiation. We have the species $\Lambda, M, R$ and $k$ (assumed
$\neq 0$). Following the algorithm outlined above we immediately
obtain the constants \cite{rindler}
\begin{equation}
\mathcal{F}_{1}=\frac{\Omega_{\Lambda}\Omega_{M}^2}{\Omega_{k}^3},\;\;\;\mathcal{F}_{2}=\frac{\Omega_{\Lambda}\Omega_{R}^3}{\Omega_{M}^4},
\label{rad12}
\end{equation}

\begin{equation}
\mathcal{F}_{3}=\frac{\Omega_{\Lambda}\Omega_{R}}{\Omega_{k}^2},\;\;\;\mathcal{F}_{4}=\frac{\Omega_{k}\Omega_{R}}{\Omega_{M}^2}.
\label{rad34}
\end{equation}

First note that the constant $\mathcal{F}_{1}$ is
\textit{inherited} from the $\Lambda$CDM model as discussed above.
Moreover, only two of the above constants are independent. For
example, $\mathcal{F}_{3}=\mathcal{F}_{1}\mathcal{F}_{4}$ and
$\mathcal{F}_{2}=\mathcal{F}_{1}\mathcal{F}_{4}^3$. In, say, the
$\Lambda, M, R$ subspace the history of a universe is obtained by
the trajectory defined by the intersection of $\mathcal{F}_{2}$
with one of the other constants and with $\Omega_{k}$ replaced by
$1-\Omega_{\Lambda}-\Omega_{M}-\Omega_{R}$ in that constant
\cite{further}.

We end our discussion of constants of the motion here by noting
that there are an infinite number of universes, of arbitrary
complexity (but containing the species $\Lambda, M$ and $k$), for
which
\begin{equation}
\frac{\Omega_{k}^3}{\Omega_{\Lambda}\Omega_{M}^2} \sim 0
\label{flatness}
\end{equation}
throughout the entire history of the universe. The fact that we
would appear to live in such a universe is a problem for some
cosmologists - the ``flatness problem". We take the view that this
``problem" arises only when too few species are considered, in the
limit a two species model with only $M$ and $k$, in order to put
the issue in perspective.

Central to the discussion given above is the system
(\ref{system}). We now discuss how this system can be used to make
a fundamental distinction between model universes.

\section{The gravitational epoch function}

In rigorous texts on general relativity, for example
\cite{O'Neill} and \cite{Sachs}, spacetime is defined (in part) as
a time-oriented manifold. As stated in \cite{O'Neill},
time-oriented is often weakened simply to time-orientable, and as
pointed out in \cite{Sachs}, the local time orientation is, with
guesswork, extrapolated to the universe as a whole. Here we take a
gravitational epoch function to be defined as a scalar field
constructed from dimensionless ratios of invariants, each
derivable from the Riemann tensor without differentiation, such
that the function is monotone throughout the history of the
universe. The purpose of an epoch function is to allow the
ordering of events without reference to any specific class of
observers or, of course, coordinates. The existence of an epoch
function, when unique, allows the rigorous time-orientation of an
entire manifold without guesswork. This is why the existence of an
epoch function is interesting.

In a conformally flat four dimensional spacetime the maximum
number of independent scalar invariants derivable from the Riemann
tensor without differentiation is four. These are usually taken to
be the Ricci scalar $R$ at degree $d = 1$ and the Ricci invariants
$r(d-1)$ for $d=2,3,4$. (Degree ($d$) here means the number of
tensors that go into the construction of the scalar. Let $R_{a b c
d}$ signify the Riemann tensor, $R_{a}^{b}$ the Ricci tensor, $R$
the Ricci scalar ($\equiv R_{a}^{a}$) and $S_{a}^{b}$ the
trace-free Ricci tensor ($ \equiv R_{a}^{b}-\frac{R}{4
}\delta_{a}^{b} $). The Ricci invariants are defined by (the
coefficients are of no physical consequence as they derrive from
the spinor forms of the invariants): $r(1) \equiv \frac{1}{2^2}
S_{a}^{b} S^{a}_{b}$, $r(2) \equiv -\frac{1}{2^3} S_{a}^{b}
S_{b}^{c} S_{c}^{a}$, and $r(3) \equiv \frac{1}{2^4} S_{a}^{b}
S_{b}^{c} S_{c}^{d}S_{d}^{a}$.  The Kretschmann invariant, for
example, is not used since $6R_{a b c d}R^{a b c d} = R^2+48 r(1)$
in the conformally flat case.) The dimension of these invariants
is $L^{-2d}$. In a Robertson - Walker spacetime only two of these
invariants are independent and in this spacetime it follows that
$r(1)^2/r(3)=12/7, \,r(1)^3/r(2)^2=3, \, R^4/r(3)=12f^2/7$ and
$(R^3/r(2))^2=3f^3$ where
\begin{equation}
f \equiv \frac{R^2}{r(1)}. \label{ratio}
\end{equation}
 It is clear that $f$ given by (\ref{ratio}), or any
monotone function thereof, is the only epoch function that can be
constructed in a Robertson - Walker spacetime. Universes can then
be fundamentally categorized as to whether or not they admit
monotone $f$.

For convenience consider $\mathcal{T} \equiv f/48$. It follows
directly that in any Robertson - Walker spacetime
\begin{equation}
\mathcal{T} = (\frac{\Omega_{k}+q-1}{\Omega_{k}-q-1})^2.
\label{epoch}
\end{equation}
Whereas $\mathcal{T}$ is defined through a bounce, its
representation in the $\Omega$ notation is not. This is of no
concern here since we are concerned with the behaviour of
$\mathcal{T}$ strictly prior to any bounce should one exist. With
the aide of (\ref{system}), applied only to the species $k$, it
also follows that
\begin{equation}
\mathcal{T}^{\;'} = \frac{4(\Omega_{k}+q-1)(-2q^2
\Omega_{k}+q^{'}(\Omega_{k}-1) )}{(\Omega_{k}-q-1)^3}.
\label{epochprime}
\end{equation}
If we now apply the restriction (\ref{barotropic}) and use the
full set of equations (\ref{system}) it follows that in addition
to (\ref{q})
\begin{equation}
q^{\;'}=\frac{9}{2}((\sum_{i}w_{i}\Omega_{i})^2-\sum_{i}w_{i}^2\Omega_{i}).
\label{qprime}
\end{equation}

Let us again consider the $\Lambda$CDM model \cite{rad}. It
follows from (\ref{epoch}) that
\begin{equation}
\mathcal{T} = (\frac{\Omega_{M}+4
\Omega_{\Lambda}}{3\Omega_{M}})^2 \label{epochdust}
\end{equation}
and from (\ref{epochprime}) and (\ref{qprime}) that
\begin{equation}
\mathcal{T}^{\;'} = 8 \Omega_{\Lambda}(\frac{\Omega_{M}+4
\Omega_{\Lambda}}{3\Omega_{M}^2}) \label{epochdustprime}
\end{equation}
and finally from (\ref{system}) applied to $M$ and $\Lambda$ that
\begin{equation}
\mathcal{T}^{\;''} = 8 \Omega_{\Lambda}(\frac{\Omega_{M}+8
\Omega_{\Lambda}}{\Omega_{M}^2}). \label{epochdustprimeprime}
\end{equation}
As a result, $\mathcal{T}$ has a global minimum value of $0$ when
\begin{equation}
\Omega_{\Lambda}=-\frac{\Omega_{M}}{4}. \label{epochdustminimum}
\end{equation}
The question then is, what $\Lambda$CDM models intersect the locus
(\ref{epochdustminimum}) and therefore fail to have $\mathcal{T}$
monotone \cite{degenerate}?

Since $\Omega_{M} > 0$ clearly all $\Lambda$CDM models with
$\Lambda > 0$ admit the epoch function (\ref{epochdust}). Now
consider $\Lambda <0$. For $k=0$ condition
(\ref{epochdustminimum}) holds at
$(\Omega_{M},\Omega_{\Lambda})=(4/3,-1/3)$. For $k \neq 0$ use
(\ref{epochdustminimum}) in (\ref{dust}) to define
\begin{equation}
\mathcal{F}_{int}=\frac{16 \Omega_{M}^3}{(3\Omega_{M}-4)^3}.
\label{dustintersection}
\end{equation}
With $k=-1$ choose a value of $\Omega_{M}$ in the range $0 <
\Omega_{M} < 4/3$ and insert this value into
(\ref{dustintersection}). The resultant $\mathcal{F}_{int}$ (which
is negative) is the constant of motion associated with the
intersection of the associated integral curve with the locus
(\ref{epochdustminimum}) with the intersection taking place at the
chosen value of $\Omega_{M}$. For $k=1$ choose a value of
$\Omega_{M}$ in the range $4/3 < \Omega_{M} < \infty$ and follow
the foregoing procedure ($\mathcal{F}_{int}$ is now positive).
However, with $k=1$, as $\Omega_{M} \rightarrow \infty$ the
constant $\mathcal{F}_{int}$ reaches the lower bound of $16/27$.
There is then the range $0 < \mathcal{F} < 16/27$ for which the
associated integral curves do not intersect the locus
(\ref{epochdustminimum}). We conclude then that for the
$\Lambda$CDM models, all models with $\Lambda > 0$ admit the epoch
function (\ref{epochdust}), but for $\Lambda < 0$ only those with
$k=1$ and $0 < \mathcal{F} < 16/27$ do. There is no evidence that
our universe lies in this latter category. Indeed, all the current
evidence would suggest that we are very far away from it.

We now make some general observations on the monotonicity of
$\mathcal{T}$, but restricted to the case $\Omega_{k}=0$
\cite{furthert}. In this case it follows from (\ref{epochprime})
that the monotonicity of $\mathcal{T}$ is closely associated with
the monotonicity of $q$. For example, if we assume that $q$ is
monotone decreasing, then we can put a limit on the maximum $w$
allowed for a monotone $\mathcal{T}$. To do this note that for
models with a Big Bang the initial value of  $q$ is $(3
\tilde{w}+1)/2$ where $\tilde{w}$ signifies the largest value of
$w$ (assumed $>-1$) for all species. It follows from
(\ref{epochprime}) then that $\mathcal{T}$ can be monotone only
for $\tilde{w}\leq 1/3$. In view of (\ref{energy}) this is a
rather surprising restriction, but one which is in accord with all
observations.

\section{Summary}

A complete set of constants of the motion has been constructed for
all FLRW models consisting of an arbitrary number of separately
conserved species, each with a constant ratio of pressure to
density. These constants of the motion are to be considered as the
fundamental characteristics of the model universes. The unique
candidate for a gravitational epoch function has been constructed
for all FLRW models. In the simplest of all models, the
$\Lambda$CDM model, it has been shown that the epoch function
exists for all models with $\Lambda > 0$, for no models with
$\Lambda = 0$, and for almost no models with $\Lambda < 0$. This
function allows the global ordering of events without reference to
any particular class of observers or, of course, coordinates.

\bigskip

\begin{acknowledgments}
It is a pleasure to thank Nicos Pelavas for comments. This work
was supported by a grant from the Natural Sciences and Engineering
Research Council of Canada.
\end{acknowledgments}


\begin{thebibliography}{}
\bibitem{nophotons}We are not considering the
propagation of photons but still use the familiar $1+z$ notation.
In particular, we can set $a>a_{o}$.
\bibitem{two}The case of two species is completely
determined by (\ref{omegasum}).
\bibitem{rad}The results when the species $R$ is also included are
somewhat similar. See \cite{further}.
\bibitem{degenerate}Since $\mathcal{T}=1/9$ for
$\Lambda=0$ we consider the associated spacetimes time-unoriented.
\end{thebibliography}
\end{document}